# AI-Educational Development Loop (AI-EDL): A Conceptual Framework to Bridge AI Capabilities with Classical Educational Theories

Ning Yu, Jie Zhang, Sandeep Mitra, Rebecca Smith, Adam Rich

State University of New York (SUNY) Brockport

**Abstract**

This study introduces the AI-Educational Development Loop (AI-EDL), a theory-driven framework that integrates classical learning theories with human-in-the-loop artificial intelligence (AI) to support reflective, iterative learning. Implemented in EduAlly, an AI-assisted platform for writing-intensive and feedback-sensitive tasks, the framework emphasizes transparency, self-regulated learning, and pedagogical oversight. A mixed-methods study was piloted at a comprehensive public university to evaluate alignment between AI-generated feedback, instructor evaluations, and student self-assessments; the impact of iterative revision on performance; and student perceptions of AI feedback. Quantitative results demonstrated statistically significant improvement between first and second attempts, with agreement between student self-evaluations and final instructor grades. Qualitative findings indicated students valued immediacy, specificity, and opportunities for growth that AI feedback provided. These findings validate AI-EDL's potential to enhance student learning outcomes through developmentally grounded, ethically aligned, and scalable AI feedback systems. The study concludes with implications for future interdisciplinary applications and refinement of AI-supported educational technologies.

## Introduction

The rapid advancement of artificial intelligence (AI) has transformed the educational landscape, offering new possibilities for scalable, personalized, and adaptive learning experiences. From intelligent tutoring systems to automated feedback tools, AI technologies have become increasingly embedded in classroom practice. Yet despite their promise, many existing AI-powered systems function primarily as technical solutions rather than pedagogically integrated tools. As recent reviews have noted (Wang et al., 2024), most AI in Education (AIED) research prioritizes automation and efficiency while overlooking the foundational theories of learning that underpin effective instructional design and student development.

This disconnect is particularly critical in domains such as teacher education, where reflective practice, feedback literacy, and intellectual character formation are central to learning. In these contexts, AI systems that operate as "black boxes", offering decontextualized answers without transparency or pedagogical alignment, fail to support the deeper cognitive and ethical goals of professional preparation. Moreover, AI tools that marginalize instructor input risk undermining the core educational principle that learning is co-constructed through human dialogue, feedback, and revision (Holstein et al., 2019; Luckin et al., 2016).

To address these challenges, this paper introduces the AI-Educational Development Loop (AI-EDL), a theory-grounded framework that models learning as an iterative, feedback-driven process supported by human-in-the-loop AI systems. AI-EDL integrates classical theories of learning, including Socratic questioning, Aristotelian virtue ethics, Hull's drive-reduction theory, Skinner's operant conditioning, and Zimmerman's model of metacognition, into a structured loop that guides learners from initial knowledge gaps through reflection, revision, and eventual mastery. In contrast to systems that offer one-shot judgments or static feedback, AI-EDL emphasizes transparency, growth, and educational intentionality in every interaction.

This framework was implemented in EduAlly, an AI-powered tutoring system designed to support undergraduate students, particularly pre-service teachers, in writing-intensive and feedback-sensitive learning tasks. EduAlly provides formative, rubric-aligned feedback after students' initial submissions, enabling a second attempt and promoting reflection and improvement. Importantly, the system allows instructors to monitor and intervene at each stage, preserving pedagogical oversight while leveraging AI's strengths in responsiveness and scalable implementation.

To examine the effectiveness and practical implications of this model, we conducted a mixed-methods pilot study at a comprehensive public university. The study investigated three core areas: (1) the alignment between AI-generated feedback, teacher evaluations, and student self-assessments; (2) the impact of iterative feedback on student performance across attempts; and (3) student perceptions of the AI feedback process. The findings inform the design and implementation of AI-assisted learning tools that not only enhance efficiency but also embody the core values of reflective, developmentally grounded education.

**Research Questions (RQ)**

This study is guided by the following research questions:

- RQ1: *To what extent do AI-generated grades align with instructor-assigned grades for student responses to content-specific questions, and to what extent do instructor-assigned grades align with student self-evaluations for student overall performance?*
- RQ2: *How does student performance change between the first and second submission attempts after receiving AI-generated feedback?*

- RQ3: *How do students perceive the impact of AI-generated feedback within the EduAlly platform?*

These questions collectively explore not only the technical performance of AI feedback, but also its alignment with human judgment, its impact on learning progression, and its reception by learners, all central to evaluating the practical and theoretical value of AI-EDL in real educational settings.

## Literature Review

As artificial intelligence becomes increasingly integrated into educational technologies, researchers have raised concerns that many AI-powered tools remain disconnected from foundational pedagogical theories. Most systems prioritize optimization and automation over reflection, agency, and developmental learning models. A recent review by Wang et al. (2024) highlights this disconnect, noting that while interest in AI for formative assessment, personalized instruction, and intelligent tutoring is growing, few systems incorporate rigorous educational theory into their design or evaluation.

The AI-Educational Development Loop (AI-EDL) addresses this gap by systematically embedding classical educational theories into the structure and function of an AI-supported feedback loop. Socratic dialogue, one of its cornerstones, frames learning as a dialogic process of questioning and reflection, promoting critical thinking rather than passive absorption (Paul & Elder, 2007). Aristotelian virtue ethics complements this view by emphasizing the cultivation of intellectual character and the formation of virtuous habits through purposeful action, a framework well-suited for iterative and goal-oriented learning (Hursthouse & Pettigrove, 2023). Hull's drive-reduction theory (1943) offers a behavioral motivation model in which students are compelled to resolve cognitive dissonance and pursue mastery, while Skinner's operant conditioning (1954) underscores how positive reinforcement can encourage meaningful learning behaviors. Zimmerman's (2002) work on metacognition and self-regulated learning further contributes to AI-EDL by highlighting students' ability to monitor and direct their own progress over time.

This theoretical foundation contrasts with many mainstream AIED systems that function as opaque, black-box technologies. Azaiz et al. (2024), in their study on GPT-4-based feedback generation for programming education, acknowledge the utility of large language models in providing detailed, adaptive feedback. However, they also stress that such systems should not operate independently of educators, as AI often lacks contextual awareness, pedagogical intent, and student-specific understanding (Azaiz et al., 2024). Similarly, Wang et al. (2024) identify "human oversight" as a critical feature in effective AI systems, particularly those targeting higher-order thinking, reflective learning, or professional preparation. In these environments, teacher input enhances interpretability, aligns AI feedback with curricular objectives, and ensures ethical decision-making, principles that AI-EDL operationalizes through its human-in-the-loop architecture.

Indeed, Holstein et al. (2019) and Luckin et al. (2016) have long argued that AI should augment rather than replace human educators. Their work supports a co-regulated model where instructors remain central to guiding student progress. EduAlly exemplifies this approach by maintaining transparency in its AI-generated feedback, allowing instructors to review student progress, evaluate revisions, and intervene when necessary. This design ensures that AI feedback supports rather than substitutes pedagogical goals, offering a mechanism for instructors to scaffold learning in meaningful and context-aware ways.

Experiential learning is also central to AI-EDL's structure. Schön's (1983) framework of reflection-in-action and reflection-on-action forms the basis for students' engagement with feedback. Rather than passively receiving grades, students in EduAlly receive immediate AI feedback, reflect on their performance, and attempt revised responses with increased depth. This practice, as emphasized by Zeichner and Liston (2014), fosters deep learning and professional readiness, especially in teacher preparation programs where iterative skill development is key. Furthermore, Wang et al. (2024) observe that few AI systems offer a well-defined theory of change or a clearly articulated learning loop. EduAlly's integration of feedback, self-assessment, and revision contributes to a rare but increasingly important category of systems designed to support continuous learning through structured interaction. It not only provides insight into the efficacy of AI-driven feedback but also advances understanding of how classical learning theories can guide the design of adaptive, reflective, and transparent educational technologies.

Despite growing recognition of the importance of theory-driven AI in education, a significant gap remains in the development of integrated frameworks that simultaneously operationalize classical learning theories, support human-in-the-loop feedback mechanisms, and promote iterative, reflective practice through real-world classroom implementation. Most existing systems either reference educational theories superficially or emphasize automation at the expense of instructor agency and student metacognition. While promising advances such as GPT-based feedback tools (Azaiz et al., 2024) and theory-aware tutoring systems (Wang et al., 2024) are emerging, they often lack a coherent developmental model that connects AI feedback cycles with deeper pedagogical constructs. Moreover, few studies systematically evaluate how such frameworks function in authentic instructional contexts, particularly in fields like teacher education, where the development of professional identity, reflective practice, scaffolded feedback, and the learning process are central. The AI-Educational Development Loop (AI-EDL) is designed to address this gap by offering a theoretically grounded, empirically validated, and experientially rich model that integrates diverse strands of educational thought into a structured AI-supported learning process. In doing so, AI-EDL contributes a novel paradigm for designing and evaluating AI-enhanced learning environments that align with long-standing educational values and emerging technological capabilities.

## Theoretical Framework: The AI-Educational Development Loop (AI-EDL)

The AI-Educational Development Loop (AI-EDL) is a conceptual framework designed to model adaptive, goal-oriented, and reflective learning in AI-powered educational environments. Figure 1 illustrates the AI-EDL Framework, demonstrating how the model integrates classical educational theories with modern artificial intelligence technologies to create a structured and cyclical learning process that emphasizes diagnosis, feedback, behavioral transformation, and achievement.

Insert Figure 1 here.

At its core, the AI-EDL framework is composed of the following sequential components:

- Knowledge Gap

The cycle begins with the identification of a knowledge gap, a cognitive dissonance or a lack of understanding that motivates the learner to engage. This stage reflects the initial need or drive for gaining knowledge.

- Learning

Learners engage with curated instructional materials designed to address the identified gap. The content is often selected or structured by an AI-powered software/agent in collaboration with a teacher.

- Trial

Students attempt to apply their learning by completing assignments, quizzes, or tasks. This active experimentation allows learners to test their understanding and application of knowledge in practice.

- Assessment

Submissions are assessed using AI-powered software that generates adaptive feedback and provisional grading. This stage provides the learner with formative insights into their performance, identifying strengths and areas for improvement.

- Reflection

Students review the feedback and engage in metacognitive processing, considering why certain answers were incorrect, how their reasoning unfolded, and what changes are needed. This critical stage simulates Socratic inquiry and fosters deeper understanding.

- Loop Continuation

If the learning goals are not fully achieved, the learner re-enters the cycle, beginning again at the Knowledge Gap stage, now informed by prior reflection and feedback. This iterative loop supports mastery learning and continuous improvement.

- Goal Achievement

Based on the refined responses (e.g., the second attempt) and overall performance, the teacher or AI determines whether the learning objectives have been achieved. Final grades or validations are issued, signaling the closure of the loop.

**Key Characteristics of AI-EDL**

AI-EDL includes the following key characteristics: (1) Personalized: The AI agent adapts content and feedback based on individual student data and performance patterns. (2) Cyclical:

Learning is not linear; it supports iterative practice, self-revision, and repeated attempts until goals are met. (3) Collaborative: Involves coordinated roles among the Student, AI Agent, and Teacher, each contributing to diagnosis, guidance, or validation. (4) Technologically enabled: Implements AI-powered analysis, feedback generation, and data tracking to enhance learning outcomes in scalable ways. (5) Theoretically grounded: Aligns with multiple educational theories, including Socratic dialogue (reflection), Aristotelian ethics (habit and virtue), Hull's drive theory (motivation), and Skinner's conditioning (reinforcement). (6) System design and implementation: A prototype of the AI-Educational Development Loop (AI-EDL) has been implemented by extending an existing system called EduAlly, a pilot AI-powered adaptive learning software developed by the research team.

Figure 2 demonstrates the three related subjects in a simple adaptive learning scenario. Three subjects are notated in Latin letters, representing Teacher (Ţ), Student (Ş), and AI Agent (Ą), which is a cloud-based intelligent system. In addition, $M$ represents instructional materials prepared by the teacher, $Q$ for questions prepared by the teacher, $K$ for answer keys prepared by the teacher, $R$ for the grading rubric prepared by the teacher, $D$ for the backend database established by the AI agent, $F$ for the adaptive feedback generated by the AI agent, $G$ for final grading provided by the teacher, and $A$ for the answers by the student.

Insert Figure 2 here.

The workflow can be broken into the following phases in the AI planning logic style:

i) Input preparation: Teacher prepares and collates the input to the AI agent. Input: $I = \{M, Q, K, R\}$, Ţ submits $I$ to Ą, $Submit$(Ţ, $I$, Ą).

ii) Data analysis and backend setup: The AI agent analyzes $I$ and initializes the backend database. $D = f_{Ą}(I)$, where $f_{Ą}$ represents the AI agent's initialization function to process and store the data. $Init(Ą, I, D)$

iii) Material rendering: The AI agent renders materials $M$ to the student, $Ą \rightarrow Ş$: $Render(Ą, M, Ş)$

iv) Student engagement: The student receives and studies the materials: $Digest\ (Ş, M)$.

v) Question rendering: The AI agent renders the questions $Q = \{q_1, q_2, \ldots, q_n\}$ to the student, $Ą \rightarrow Ş$: $Render(Ą, Q, Ş)$

vi) Answer submission: The student answers the questions, $Answer\ (Ş, Q, A_{Ş})$ and submits them to the AI agent, $Submit\ (Ş, A_{Ş}, Ą)$, where $A_{Ş} = \{a_1, a_2, \ldots, a_n\}$ represents the answer from the student.

vii) Feedback generation: The AI agent analyzes answers and generates adaptive feedback and the grade: $(F, G_{Ą}) = g(A_{Ş}, K, R)$, where $g$ represents the function of AI agent to generate adaptive feedback $F$ and the grade $G_{Ą}$ according to the input $A_{Ş}$, the answer key $K$, and grading rubric $R$.

viii) Feedback rendering: The AI agent renders the feedback and initial grading to the student, $Render(Ą, \{F, G_{Ą}\}, Ş)$, where $G_{Ą}$ is the grade generated by the AI agent.

ix) Feedback review: The student reviews and digests the feedback and the grade, and implicitly reflects on the learning.

$Review\ (Ş, \{F, G_{Ą}\}), Digest\ (Ş, \{F, G_{Ą}\}), Reflect(Ş, \{M, Q, A_{Ş}, F, G_{Ą}\})$.

x) Second attempt: The student revises and submits the revised answers to the AI agent.

$Revise(Ş, A_{Ş}, A'_{Ş}), Submit\ (Ş, A'_{Ş}, Ą)$, where $A'_{Ş}$ is the revised answer, $A'_{Ş} = \{a'_1, a'_2, \ldots, a'_n, a'_{n+1}\}$, $a'_{n+1}$ represents explicit self-reflection on the learning.

xi) Attempt forwarding: The AI agent forwards $A'_{Ş}$ to the teacher for review,

$Forward(Ą, A'_{Ş}, Ţ)$.

xii) Teacher grading: The teacher reviews the answers and provides the final grade.

$G = g_{Ţ}(A'_{Ş}, K, R)$, where $g_{Ţ}$ represents the teacher's grading function.

xiii) Grading submission: The teacher submits $G$ to the AI agent, $Submit(Ţ, G, Ą)$.

xiv) Final feedback rendering: The AI agent renders $G$ to the student, $Render(Ą, G, Ş)$.

**Mapping Between AI-EDL and Its Implementation**

EduAlly follows a well-defined adaptive learning workflow that aligns closely with the four stages of the AI-EDL framework. Table 1 outlines the detailed implementation of AI-EDL, delineating each step with its corresponding prototype phase and participants involved (e.g., Teacher, AI agent, Student).

Insert Table 1 here.

**Theoretical Alignments for AI-EDL**

Furthermore, each phase of the prototype implementation (as described in the AI-EDL with Student–AI Agent–Teacher interactions in Table 1) is aligned with the four major educational theories or philosophical traditions, explained in Table 2:

1. Socratic dialogical method (emphasizing reflection and questioning)
2. Aristotelian virtue ethics (emphasizing habituated excellence and purpose)
3. Hull's drive-reduction theory (emphasizing motivation via resolution of cognitive tension)
4. Skinner's operant conditioning (emphasizing learning through reinforcement and feedback)

Insert Table 2 here.

## Methods

This study employed a mixed-methods approach, integrating quantitative and qualitative analyses to examine: (1) the differences between AI-generated grades, teacher-assigned grades, and student self-evaluations; (2) changes in student performance between initial and revised submissions (Attempt I vs. Attempt II), following AI-generated feedback; and (3) student perceptions regarding the impact of AI-generated feedback within the EduAlly platform.

### Participants and Setting

After receiving ethical approval, this pilot study was conducted at a mid-size comprehensive public university located in the northeastern United States. Participants consisted of undergraduate and graduate students enrolled in special education courses taught by the second author. All participants provided informed consent prior to participation in this research.

### Statistical Analyses

Descriptive statistics were used to calculate the means and standard deviations of student performance graded by the AI, the teacher, and student self-evaluations. To compare these grades across these three evaluators, Wilcoxon signed-rank tests were conducted. In addition, Spearman's rank-order correlation analyses were employed to explore the relationships between student level and key assessment variables, including AI-generated grades, teacher-assigned grades, and student self-evaluations. Qualitative analysis was also conducted to investigate student perceptions regarding the impact of AI-generated feedback within the EduAlly platform.

## Results and Findings

### Quantitative Results

Descriptive statistics were employed to calculate the means and standard deviations of student performance graded by the AI, teacher, and student self-evaluations, which were reported in Table 3. An alternative grading system was implemented using a rubric with three performance levels (i.e., *Satisfactory* = 2, *Improvement Needed* = 1, and *Not Assessable* = 0). The AI grade and teacher grade for both Attempt I and Attempt II assessed individual responses to content-specific questions, excluding self-evaluation and self-reflection items. Technical difficulties with the platform prevented some students from submitting some of their responses, resulting in missing data for AI-generated feedback and AI grades. Consequently, in Attempt I, the number of AI-graded questions ($n$ = 918) was smaller than that of the teacher-graded questions ($n$ = 962).

The results of the descriptive statistics indicated that in Attempt I, AI graded 918 submissions and provided feedback, with a mean score of 1.40 (SD = 0.73), ranging from 0 (no feedback) to 2 (satisfactory). In comparison, the teacher graded 962 submissions, with a higher average score of 1.56 (SD = 0.73), also ranging from 0 to 2. For Attempt II, the teacher graded 963 submissions, showing an increase in the mean score to 1.77 (SD = 0.57), ranging from 0–2.

In contrast, the student self-evaluation and teacher final grade reflected holistic assessments of overall student performance. Not all participating students completed the specific question on self-evaluation in Attempt II; therefore, teacher final grades ($n$ = 114) exceeded student self-evaluations ($n$ = 74). Student self-evaluations reported a notably high average score of 1.99 (SD = 0.12), ranging from 1 (improvement needed) to 2 (satisfactory). After reviewing both attempts, the teacher provided a final holistic grade, with the mean final grade of 1.89 (SD = 0.37), ranging from 0 (not assessable) to 2 (satisfactory).

Insert Table 3 here.

Wilcoxon signed-rank tests were conducted to compare students' grades across three evaluators (i.e., AI, the teacher, and students). The results revealed that for students' Attempt I submission, a statistically significant difference in grades was assigned by the teacher and those assigned by AI ($Z = 11.14$, $p < .001$), indicating that the teacher assigned significantly higher grades than AI on average. Further analysis of the difference revealed that AI rigidly adhered to the predetermined answer keys, while the teacher demonstrated flexibility when assessing student answers. For instance, when grading students' responses to the question "Who are the required members for this team?", the AI assigned an "I" (Improvement needed) and asked for additional elaboration when student responses deviated from the exact answer key provided. Conversely, the teacher assigned an "S" (Satisfactory) for responses that were correct despite not matching the answer key exactly.

There was a statistically significant difference in the teacher-assigned grades for Attempt I and Attempt II submissions ($Z = 10.56$, $p < .001$), indicating that students received significantly higher grades on their resubmissions (Attempt II) compared to their initial submissions (Attempt I). In addition, a statistically significant difference was found between teacher-assigned grades for Attempt II and AI-assigned grades for Attempt I ($Z = 15.12$, $p < .001$), indicating that teacher-assigned grades for student resubmissions (Attempt II) were significantly higher than AI-assigned grades for initial submissions (Attempt I).

In contrast, no statistically significant difference was found between students' self-evaluations and the teacher's final grades ($p > .05$), suggesting a high level of agreement between students' self-assessments and the teacher's final holistic evaluations. Table 4 indicates the Wilcoxon Signed-Rank test results.

Insert Table 4 here.

Spearman's rank-order correlation tests were conducted to examine the relationship between student level (coded as *Undergraduate* = 1, *Graduate* = 2) and key assessment variables, including AI-generated grades, teacher-assigned grades, and student self-evaluations. To minimize potential grading bias, both AI and the teacher remained blinded to the students' names or academic status (i.e., undergraduate or graduate level) throughout the assessment process. The results of the analyses revealed a statistically significant positive correlation between student level and AI-assigned grades (Spearman's $\rho = .08$, $p = .013$, $n = 918$), indicating that, compared to undergraduate students, graduate students were slightly more likely to receive higher grades from AI. In addition, a statistically significant positive correlation was found between student level and teacher-assigned grades for student Attempt I submission (Spearman's $\rho = .11$, $p < .001$, $n = 962$), as well as for student Attempt II resubmission (Spearman's $\rho = .10$, $p = .001$, $n = 963$). These results indicate that graduate students were more likely to receive higher teacher-assigned grades than undergraduate students across both attempts.

In contrast, no statistically significant correlations were found between student level and either student self-evaluation scores or teacher final grades ($p > .05$), suggesting that overall performance evaluation by students or the teacher was not influenced by student academic level. Table 5 illustrates the correlations between student level and key assessment variables.

Insert Table 5 here.

## Qualitative Findings

Open coding, a process involving reading and re-reading student self-reflections, initial coding, refining codes, and creating thematic categories (Saldaña, 2009), was implemented to analyze qualitative data. This approach aimed to center pre-service teachers' own words to amplify their voices, responding to the question, *How would you describe your experience with the AI feedback system?* (Clark & Creswell, 2015; Rak et al., 2021).

### ***Balance and Constructiveness of the Feedback***

Students valued that the AI feedback included both strengths and areas for improvement. One student stated, “I liked that it focused on both strengths and areas for growth, which made the feedback feel balanced.” Another student shared a similar thought, “It also highlighted what I did well so I was able to recognize my strengths and improve…”

***Clarity and Specificity of the Feedback***

Students appreciated that the feedback provided by AI was easy to understand and clearly presented. "The feedback was clear and specific, and told me specifically what needed improvement and how to better my responses," as one student exclaimed, while another student commented, "The AI feedback was very beneficial and gave me explanations on why I need improvement or why I got satisfactory."

***Encouragement to Elaborate and Deepen Responses***

Many students noted that AI encouraged them to expand on their answers and provide more detail. For example, one student shared, “The AI feedback system not only praised me for the correct response but also encouraged me to elaborate more and be more detailed in my responses the next time.” Another student agreed, “It helped me to be able to elaborate on my responses in order to improve them while also telling me the areas of my answers that were well done.”

***Immediacy and Efficiency of Feedback Delivery***

Many students reported real-time feedback as a significant benefit, allowing for immediate reflection and revision. "I love receiving instant feedback on my work and this provided just that," as one student remarked, and another student expressed a similar sentiment, “The AI feedback system was helpful as it provided me with specific and immediate feedback on my responses, whether they needed improvement or not.”

***Perceived Helpfulness of the Feedback***

Students generally found the AI feedback easy to understand and helpful in improving their responses and understanding, as students reflected: “It was very helpful and easy to understand. I love receiving instant feedback on my work and this provided just that.” “The feedback was clear, helpful, and provided useful insights for improving my work.”

***Positive Learning Experience and Growth***

Many students appreciated the opportunity to revise and improve their work based on iterative feedback, which helped deepen their understanding and learning process. One student admitted, "I really liked that I got a second attempt to fix my mistakes. I think learning from my mistakes is the best way to learn." Similar statements were made by other peers: “I found the AI feedback to be very helpful due to its feedback offering hints to how a response could go more in depth.”

***Varied Emotional Response to AI Integration in Learning***

Students expressed a range of emotions—from appreciation and curiosity to frustration and anxiety—highlighting the affective dimension of AI integration in education and showing the emotional complexity when interacting with generative AI. Most students expressed appreciation for the integration of AI in this learning process and trust in AI’s feedback, and shared positive overall sentiments. For instance, “I did really enjoy using this platform for the assignment.” “I would be happy to see this platform used for future assignments.”

Some students expressed hesitation, concerns, skepticism, or distrust in AI’s accuracy or appropriateness for grading, such as “This was a very frustrating experience.” “I don’t think AI is always accurate and don’t prefer its use in grading.”

Other students held mixed feelings toward using AI in the course: "Overall, I have a mixed opinion of the AI feedback system." “I think the system still needs refining before being used as a sole grader but is a generally helpful tool.”

***Concerns About the Clarity and Specificity of AI Feedback***

Students noted that AI feedback lacked specificity and clarity in recognizing context, especially when it suggested to “elaborate further.” One student mentioned, "I do feel that on one or two of the answers what the AI was looking for and what I had written down were the same thing, however the AI didn't recognize it within text." A similar observation was made: “There was insufficient detail surrounding the expectations… and while my initial answer was succinct and objectively correct, I was prompted for more detail.”

***Technical Challenges and User Experience Issues***

Several students encountered technical issues such as unsaved answers, system glitches, and confusion about saving progress: "... the program kept telling me that my answers were saved but then the page reloaded and I lost all of my progress." “I had a hard time because I did not read the directions fully and forgot to save my first answers which did not allow to receive any feedback.” “Although it said my answers would be saved, when I returned to complete and turn in my work, my answers were not saved … I had to redo all of my answers in a Word document and then copy and paste my answers because I didn’t trust this program after that.”

***Suggestions for Making Feedback Accessible for Future Reference***

Several students experienced an inability to revisit feedback. "I wish there was a way to save the feedback for future reference." “I wish there was a way to go back and view the feedback once the second attempt had started.”

These findings, through analyzing student perceptions on the use of AI feedback, provide valuable insights for future research and implementation. The positive student responses inspire us to continue this research. Based on student feedback, several improvements will be implemented: (1) refinement of content instruction and answer keys to enable more specific AI-generated feedback, (2) enhancement of the EduAlly platform to address technical issues, and (3) development of an accessible feedback archive for student future reference.

## Visual Analysis

### *Grading Patterns*

Figure 3 demonstrates the comparison of AI-generated feedback grades (in blue) and teacher-assigned grades (in red) for Attempt I, organized by coded student ID, question number, and semester. Matching grade points appear as overlapping red and blue markers, while discrepancies are indicated by visible separations between the two colors. The figure reveals a significant number of matched cases, where AI and teacher evaluations align, as well as a notable portion of mismatches, indicating divergence in grading judgments. This visual spectrum highlights both the potential reliability and the variability of AI feedback in initial attempts.

Insert Figure 3 here.

Figure 4 demonstrates the isolated visualization of discrepancies between AI-generated feedback and teacher-assigned grades for Attempt I, filtered to include only instances where the two evaluations differ. The majority of cases show higher teacher grades (in red) and lower AI grades (in blue), suggesting a conservative tendency in the AI's evaluation algorithm. One notable exception appears where the AI assigned a higher grade than the teacher's. Upon review, the teacher noted that this instance involved an error stemming from ambiguous learning materials, highlighting the importance of human oversight in interpreting content-sensitive responses.

Insert Figure 4 here.

The statistics of grading alignment and data completeness are summarized in Tables 6 and 7, respectively. As shown in Table 6, out of 782 responses where both AI and teacher grades were available, 654 (83.63%) had identical grades, while 128 responses (16.37%) showed discrepancies. This high level of agreement suggests strong general alignment between AI-generated and instructor-assigned evaluations.

Insert Table 6 here.

Figure 5 presents a side-by-side visualization of teacher-assigned grades in Attempt I (in red) and Attempt II (in blue) across all submissions. If no performance change occurred, the red and blue points would overlap perfectly. However, the prominent visibility of blue points above red points indicates widespread improvement in student performance following AI-supported feedback and revision.

Insert Figure 5 here.

As detailed in Table 7, when comparing teacher-assigned grades in Attempt I and Attempt II, out of 962 responses where teacher grades for both Attempt I and II were available, approximately 84.51% of responses showed no change ($n$ = 813), 15.18% demonstrated improvement ($n$ = 146), and only 0.31% declined ($n$ = 3). The decline was primarily due to students skipping the second attempt. These patterns highlight the pedagogical impact of the AI-EDL framework in fostering reflective learning and iterative improvement.

Figure 6, in combination with Table 8, illustrates the relationship between AI-generated feedback on Attempt I and teacher-assigned grades on Attempt II. Ideally, if students followed AI

guidance accurately and the AI feedback was pedagogically aligned, the blue (AI) and red (teacher) points would overlap. However, the prevalence of red points above blue, out of 783 responses where both AI grades for Attempt 1 and teacher grades for both Attempt II were available, representing 26.18% of the matched records ($n = 205$), indicates meaningful student improvement following feedback. Only 0.26% of cases show AI assigning higher grades than instructors ($n = 2$), and 73.56% show perfect agreement ($n = 576$). These patterns suggest that AI feedback not only aligns well with instructor expectations but also serves as an effective scaffold for learning and revision under the AI-EDL framework.

Insert Figure 6 here.

Insert Table 8 here.

## Discussion

The findings from this study offer strong support for the AI-Educational Development Loop (AI-EDL) as a theory-driven model for integrating artificial intelligence into reflective, instructor-supported learning environments. By embedding classical educational theories into an iterative, transparent feedback loop, EduAlly not only facilitated measurable improvement in student performance but also supported student engagement and self-regulated learning, key components of the AI-EDL framework.

First, the observed improvement in student scores between Attempt I and Attempt II confirms the value of iterative feedback in supporting learning gains. This aligns with prior research on formative assessment and revision-based learning (Zeichner & Liston, 2014; Wang et al., 2024), while providing a structured, AI-supported mechanism for implementing those practices at scale. The data suggest that timely, rubric-informed AI feedback can meaningfully guide students toward more accurate, complete, and thoughtful responses. These results empirically reinforce the theoretical proposition of AI-EDL: that meaningful learning emerges from cycles of dissonance, reflection, and revision.

Second, the high degree of alignment between student self-evaluations and final teacher-assigned grades indicates that students not only received feedback but also internalized it. This suggests that AI feedback prompted metacognitive awareness and judgment, key traits in Zimmerman's model of self-regulated learning. Importantly, the EduAlly platform was designed to support such outcomes by encouraging students to reflect on feedback before revising their responses. This reinforces AI-EDL's distinction from one-shot automated grading systems, emphasizing growth and developmental feedback rather than judgment and finality.

Third, students' qualitative reflections provided valuable insights into how AI feedback is experienced affectively and cognitively. Many students appreciated the clarity, specificity, and immediacy of the feedback, echoing Azaiz et al.'s (2024) findings that large language models can generate feedback with human-like qualities. At the same time, students expressed concerns about perceived harshness, errors in logic, and the emotional impact of receiving AI responses. These comments highlight the continued need for human-in-the-loop models, as advocated by Holstein et al. (2019) and Luckin et al. (2016), where instructors can validate, edit, or contextualize AI outputs. EduAlly's sandbox structure supports this role, preserving instructor agency and mitigating the risks of over-reliance on opaque AI systems.

Collectively, the results validate the practical utility of AI-EDL while also underscoring several critical design principles for future AI-assisted learning environments: transparency, instructor oversight, structured revision opportunities, and emotionally aware feedback. These principles are particularly important in teacher preparation and other professional education contexts, where character development, reflective practice, and ethical reasoning are core learning outcomes.

Nonetheless, several limitations must be acknowledged. First, the pilot study was conducted at a single institution with a relatively small sample size, which may constrain the generalizability of the findings. Second, student self-evaluation data were unevenly distributed, potentially biasing the analysis of metacognitive development. Finally, while EduAlly was designed for writing-intensive tasks, further investigation is needed to assess how the AI-EDL framework performs across diverse disciplines, especially in more quantitative or collaborative contexts. Addressing these limitations will be essential in future research to refine the model and strengthen its applicability across broader educational settings.

Overall, this study contributes to the growing literature on AI in education by offering a theoretically grounded, empirically tested model that bridges AI capabilities with foundational pedagogical values. The AI-EDL framework advances beyond existing black-box or behaviorist AI applications by integrating classical theories of reflection, motivation, and character into a learning process that is transparent, developmental, and human-centered.

## Conclusion

This study introduces and evaluates the AI-Educational Development Loop (AI-EDL), a theoretically grounded framework for integrating artificial intelligence into reflective, feedback-driven learning environments. Through the implementation of EduAlly, an AI-assisted platform that facilitates formative feedback, iterative revision, and instructor oversight, this work demonstrates the feasibility and pedagogical value of aligning AI systems with classical educational theories. Students not only improved their performance across attempts but also engaged in meaningful reflection and self-assessment, affirming the developmental potential of AI-EDL's cyclical design.

By incorporating principles from Socratic dialogue, drive-reduction theory, operant conditioning, and self-regulated learning, AI-EDL offers a rare synthesis of classical learning theory with emerging AI technologies. Unlike many current systems that prioritize automation over pedagogy, EduAlly exemplifies how human-in-the-loop models can preserve instructor agency, support student metacognition, and promote transparent, growth-oriented learning. The study's findings support the viability of this approach and suggest broader implications for designing AI tools that are not only effective but also ethically and educationally responsible.

Future work will extend this model to new disciplinary contexts, expand the scope of instructor engagement, and refine the AI feedback mechanisms based on student emotional and cognitive responses. As AI continues to reshape educational practice, frameworks like AI-EDL are essential to ensuring that innovation remains firmly rooted in sound pedagogical principles and responsive to the needs of learners and educators alike.

**Figure 1**

*The Illustration of the AI-EDL Framework*

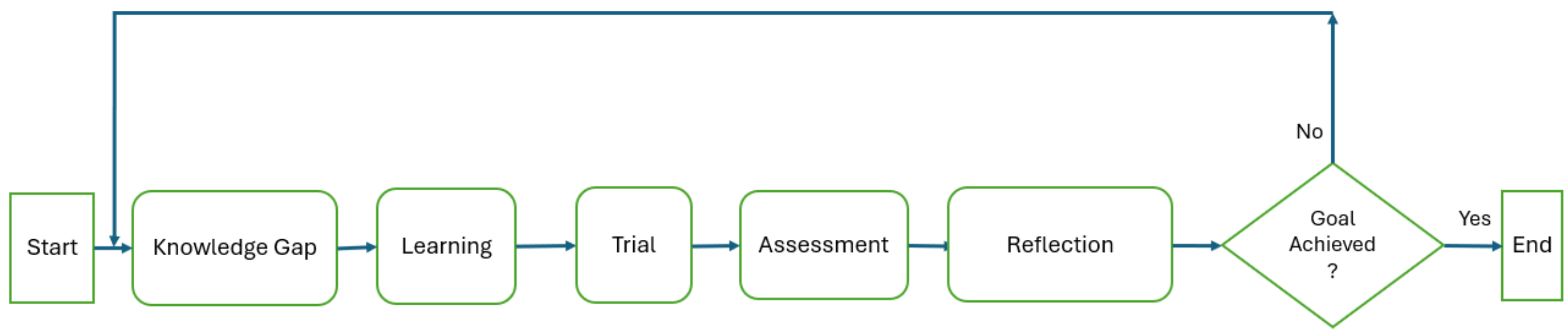

**Figure 2**

*Three Related Subjects in a Simple Adaptive Learning Scenario*

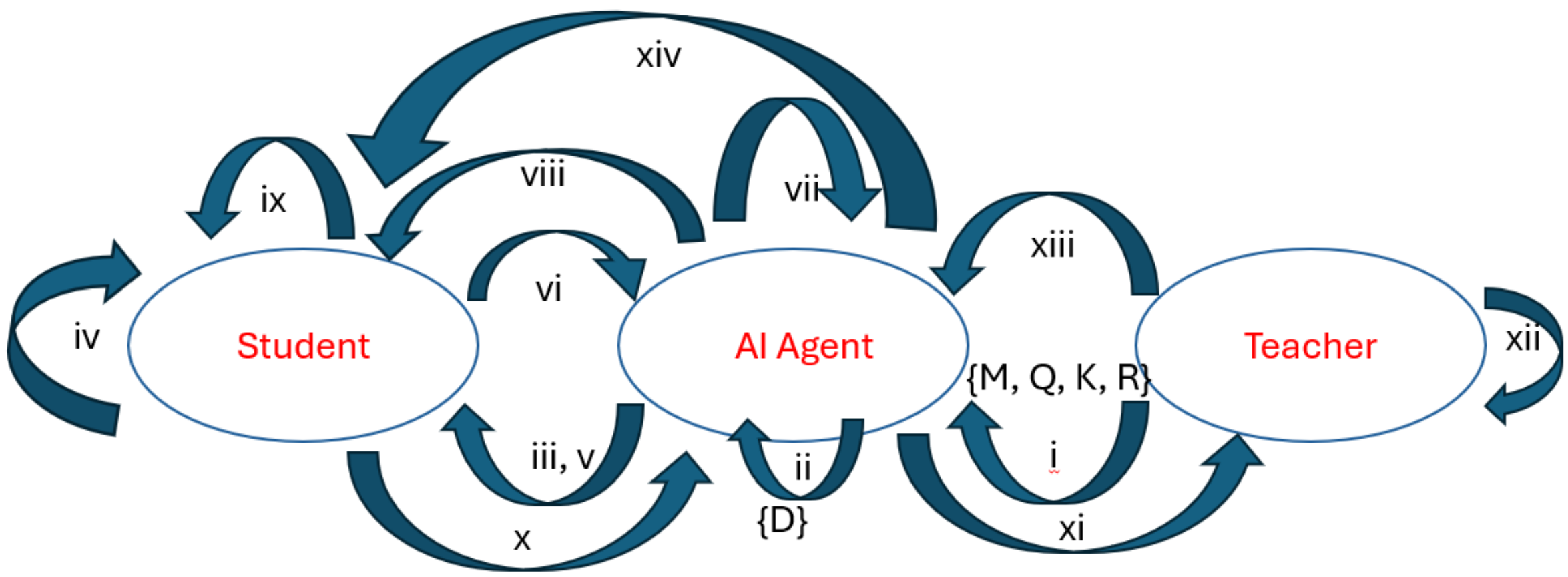

**Figure 3**

*Overlay of AI Grades (in blue) vs Teacher Grades (in red) in Attempt I*

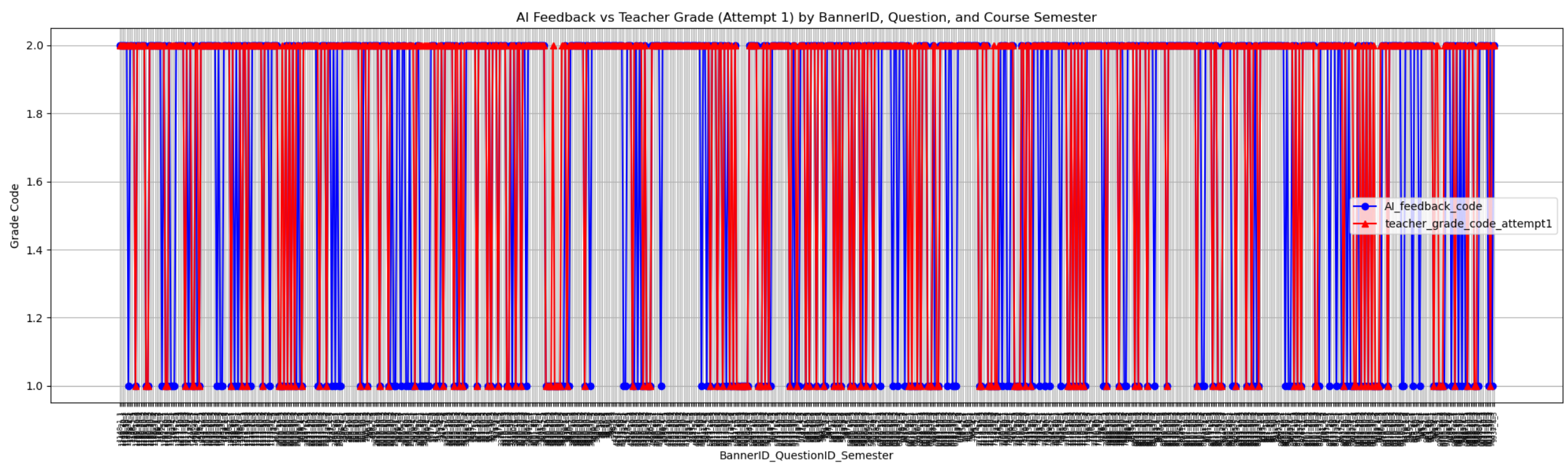

**Figure 4**

*Isolated Discrepancies Between AI Grades (in red) and Teacher Grades (in blue) for Attempt I*

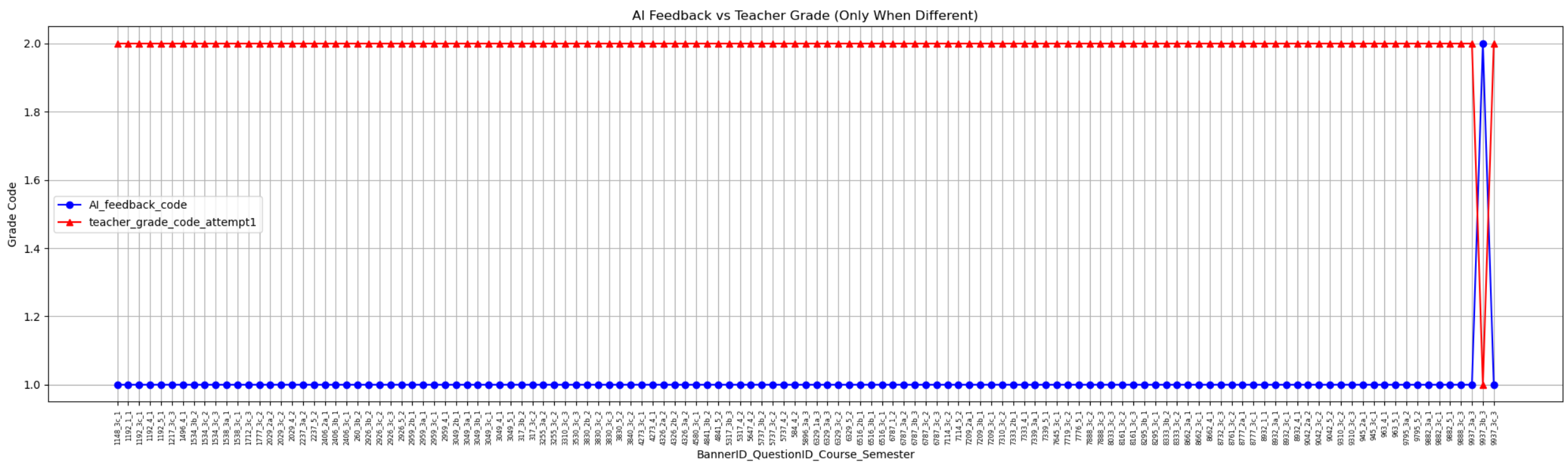

## Figure 5

*Comparison of Teacher's Grades Between Attempt I (in red) and Attempt II (in blue)*

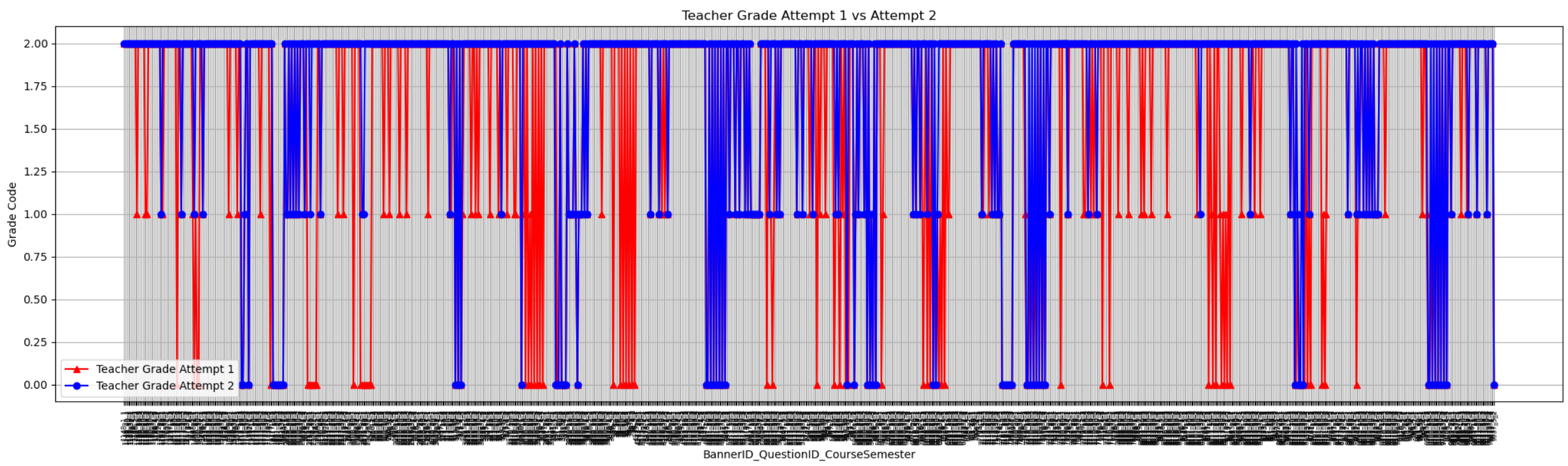

## Figure 6

*Comparison of AI Grades for Attempt I (in blue) and Teacher Grades for Attempt 2 (in red)*

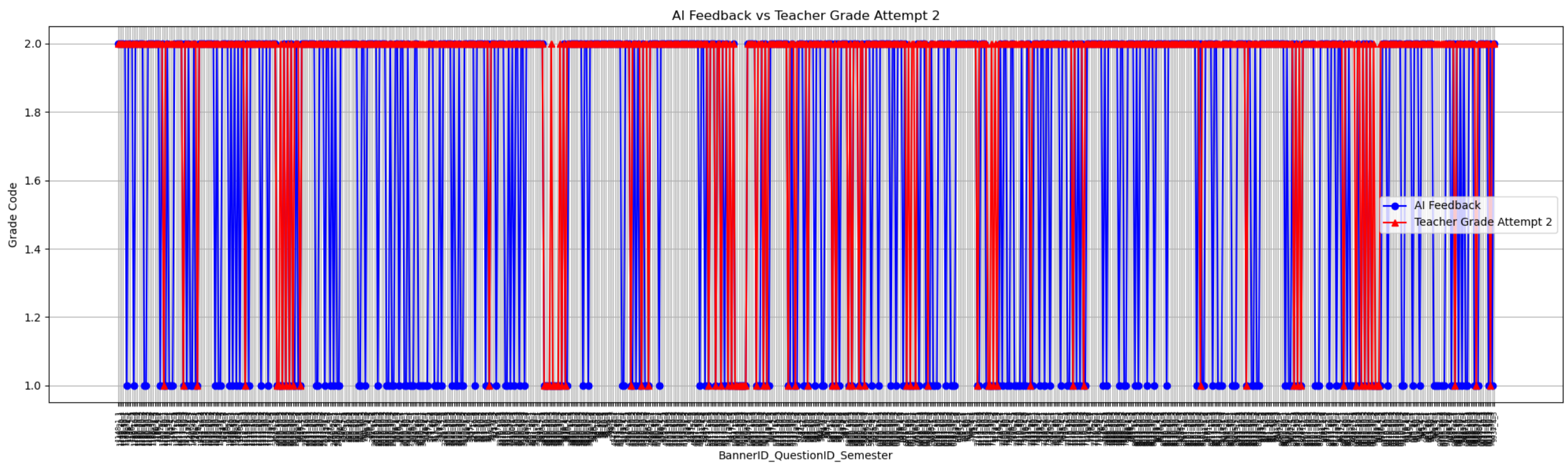

**Table 1**

*Mapping Between AI-EDL and its Implementation*

| AI-EDL Step | Corresponding Prototype Phase(s) | Participants Involved | Explanation |
|---|---|---|---|
| Start | Phases i, ii | Teacher, AI agent | Represents the activation or launch of the learning activity/system (EduAlly). |
| Knowledge Gap | Phase iv: Student engagement | Student | The student encounters material and identifies gaps in understanding. |
| Learning | Phase iii: AI renders M and Q → Student<br>Phases iv & v: Student studies M and Q | AI Agent, Student | AI provides materials; the student studies them. |
| | Phase ix: Student reviews and digests feedback (F) | | |
| Trial | Phase vi: Student submits initial answers ($A_s$) | Student, AI Agent | The student attempts to answer questions, applying what was learned. |
| | Phase x: Revised answers submitted | | |
| Assessment | Phase vii: AI evaluates answers and generates feedback (F, $G_a$)<br>Phase viii: Feedback rendered to Student | AI Agent, Student | AI provides adaptive feedback and provisional grading. |
| | Phases xii: Teacher grades final answers. | AI Agent, Teacher | |
| Reflection | Phase ix: Student reflects on learning | Student | The student internalizes feedback and considers how to revise responses. |
| Goal Achievement | Phase xiii–xiv: Final grade sent to student | Teacher, AI Agent, Student | The teacher performs final grading and feedback. |
| End | – | – | Marks the conclusion of the learning loop once learning objectives are met. |

**Table 2**

*Theoretical Alignments for AI-EDL*

| Prototype Phase | Description | Socratic Method | Aristotelian Ethics | Hull's Drive-Reduction Theory | Skinner's Operant Conditioning |
|---|---|---|---|---|---|
| Phase I (Teacher prepares input {M, Q, K, R}) | Teacher defines learning objectives and grading criteria | – | Establishes the telos (purpose) of learning; clear goal setting | – | Sets up the reinforcement schema (e.g., grading rubric) |
| Phase ii (AI agent analyzes input and initializes backend {D}) | AI processes and organizes the instructional data | – | – | – | Prepares the system for automated, consistent feedback |
| Phases iii & v (AI renders M and Q to Student) | Learner receives material and questions | Begins cognitive engagement for questioning and reflection | Initiates intellectual virtue through exposure and challenge | Stimulates knowledge-driven behavior (recognition of knowledge gap) | Initiates conditions for trial behavior to be reinforced later |
| Phase iv (Student studies M and Q) | Learner processes material | Internal Socratic dialogue begins as the learner questions and digests material | Encourages habit of self-study and discipline | Rising drive to resolve lack of knowledge | Prepares behavioral basis for reinforcement through trial |
| Phase vi (Student submits answers $A_s$) | Learner performs and submits work | Acts as a hypothesis test in learning through dialectic | Practice toward intellectual excellence via application | Learner acts under cognitive tension (drive) | Initiates operant behavior to be shaped by coming feedback |
| Phase vii (AI generates feedback F & grade $G_a$) | AI evaluates and generates feedback | Functions as a dialectical counterpoint prompting reevaluation | Feedback guides formation of virtuous habits through correction | Partial drive reduction via performance feedback | Primary reinforcement stage: feedback shapes future behavior |
| Phase viii (AI sends F and $G_a$ to Student) | Learner receives AI-generated formative assessment | Triggers reflection and deeper questioning | Reinforces disciplined learning through formative feedback | Reduces uncertainty, partially satisfying drive | Reinforcement applied; behavior strengthened or corrected |
| Phase ix (Student digests F and $G_a$) | Student reflects on performance and feedback | Central to Socratic method: learner evaluates assumptions, reasoning | Cultivates practical wisdom (phronesis) via self-correction | Reinforces or restarts motivation based on satisfaction level | Critical reinforcement checkpoint before second attempt |
| Phase x (Student submits revised answers $A'_s$) | Learner makes a second, improved attempt | Reflective action leads to refined thinking, a Socratic ideal | Displays virtue of persistence and refinement | Reactivation of drive to close remaining knowledge gap | Operant repetition: revised behavior for improved reinforcement |
| Phase xi (AI forwards $A'_s$ to Teacher) | Revised submission is passed to expert for judgment | – | Transition from practice to mastery | Final push toward goal satisfaction | Final round of behavior-to-consequence loop |
| Phase xii (Teacher evaluates and finalizes grade G) | Expert-based summative evaluation | Marks closure of dialectical process through authoritative insight | Fulfillment of telos (purpose-driven evaluation) | Complete drive reduction, knowledge verified and stabilized | Strongest reinforcement signal, formal performance consequence |
| Phase xiii–xiv (Final grade G sent to Student) | Final feedback delivered to student | Outcome prompts potential new questioning and future inquiry | Completion of virtuous cycle; readiness for future endeavors | Feedback resolves cognitive tension fully | Behavior either consolidated or revisited in next learning loop |

**Table 3**

*Descriptive Statistics*

| | N | Mean | Std. Deviation | Minimum | Maximum |
|---|---|---|---|---|---|
| | | Attempt I | | | |
| AI Grade | 918 | 1.40 | .73 | 0 | 2 |
| Teacher Grade | 962 | 1.56 | .73 | 0 | 2 |
| | | Attempt II | | | |
| Teacher Grade | 963 | 1.77 | .57 | 0 | 2 |
| Student Self-eval | 74 | 1.99 | .12 | 1 | 2 |
| Teacher Final Grade | 114 | 1.89 | .37 | 0 | 2 |

**Table 4**

*Wilcoxon Signed-Rank Test Results*

| | Attempt I: Teacher Grade - AI Grade | Teacher Grade: Attempt II- Attempt I | Teacher Grade Attempt II - AI Grade Attempt I | Attempt II: Teacher Final Grade - Student Self-Eval |
|---|---|---|---|---|
| Z | 11.14 | 10.56 | 15.12 | 1.000 |
| Asymp. Sig. (2-tailed) | <.001 | <.001 | <.001 | .317 |

**Table 5**

*Correlations Between Student Level and Key Assessment Variables*

| Variable | Spearman's ρ (rho) | *p*-value | N |
|---|---|---|---|
| AI Grade – Attempt I | 0.08* | .013 | 918 |
| Teacher Grade – Attempt I | 0.11** | <.001 | 962 |
| Teacher Grade – Attempt II | 0.10** | .001 | 963 |
| Student Self-Evaluation | 0.13 | .281 | 74 |
| Teacher Final Grade | 0.10 | .307 | 114 |

Note: ** indicates that $p < .01$,
* indicates that $p < .05$.

**Table 6**

*Statistics of AI-Generated Grades and Teacher-Assigned Grades Comparison*

| AI and Teacher Grades | N | Percentage (%) |
|---|---|---|
| Rows with both AI & Teacher Grades present | 782 | 100.00% |
| Same Grades | 654 | 83.63% |
| Different Grades | 128 | 16.37% |

**Table 7**

*Statistics of Teacher-Assigned Grades for Attempts I & II*

| Teacher Grades for Attempt I vs. Attempt II | N | Percentage (%) |
|---|---|---|
| Rows with both Teacher Grades for Attempt I & II present | 962 | 100.00% |
| Attempt I > Attempt II | 3 | 0.31% |
| Attempt I < Attempt II | 146 | 15.18% |
| Attempt I = Attempt II | 813 | 84.51% |

**Table 8**

*Statistics of AI Grades for Attempt I and Teacher Grades for Attempt II*

| AI Grades for Attempt I vs. Teacher Grades for Attempt II | N | Percentage (%) |
|---|---|---|
| Rows with both AI Grades for Attempt I & Teacher Grades for Attempt II present | 783 | 100.00% |
| AI Grades Attempt I > Teacher Grades Attempt II | 2 | 0.26% |
| AI Grades Attempt I < Teacher Grades Attempt II | 205 | 26.18% |
| AI Grades Attempt I = Teacher Grades Attempt II | 576 | 73.56% |